\documentclass[aip,reprint]{revtex4-1}
\usepackage{graphics}

\draft 

\begin{document}


\title{An effective method to estimate composition amplitude of spinodal decomposition for atom probe tomography validated by phase field simulations} 

\author{Wei Xiong}
\email[]{wxiong@yahoo.com}
\affiliation{Department of Materials Science and Engineering, KTH Royal Institute of Technology, SE-100 44 Stockholm, Sweden}
\author{John {\AA}gren}
\affiliation{Department of Materials Science and Engineering, KTH Royal Institute of Technology, SE-100 44 Stockholm, Sweden}
\author{Jing Zhou}
\affiliation{Department of Materials Science and Engineering, KTH Royal Institute of Technology, SE-100 44 Stockholm, Sweden}


\begin{abstract}
Reasonable evaluation of composition amplitude in spinodal decomposition is a challenge to microanalysis of atom probe tomography, especially at early stages when phase separation is not prominent. This impedes quantitative analysis of spinodal structure in atom probe tomography as well as comparison with simulated results from phase field simulations. We hereby report an effective method to estimate the composition amplitude by constructing an amplitude density spectrum. This method can sensitively determine the composition amplitude at early stages. In particular, it substantially bridges experimental and simulation techniques comprising both discrete and continuum data in the study of spinodal decomposition. Moreover, it was found that the commonly adopted Langer-Bar-on-Miller method for atom probe analysis underestimates the composition amplitude of spinodal decomposition. Case studies have been performed on the Fe-Cr binary alloys.
\end{abstract}


\maketitle 

\section{Introduction}
In the study of phase separation, atom probe tomography (APT) is a preferential technique to determine segregation of solute atoms due to its atomic-scale resolution of the 3-dimensional (3D) distribution of atoms \cite{Marquis2010, Miller1996, Seidman2007}. In order to elucidate mechanism of phase transformation in materials, Monte Carlo (MC) technique is usually considered as a simulation tool to provide data for comparison with APT, since both techniques generate the discrete information of the solute atoms as the direct outputs. However, MC is limited by its time and length scale, and thus difficult to predict microstructure evolution with the experimental time scale. Alternatively, phase field method (PFM) is often employed, which is based on a continuum approximation developed by Hillert \cite{Hillert1961} and Cahn and Hilliard \cite{Cahn1958}. Quantitative comparison for spinodal structure between APT and PFM depends on reasonable estimation of the composition amplitude and wavelength. Additionally, one may easily envisage that estimation of composition amplitude is more important than wavelength at the early stages of spinodal decomposition, when spinodal decomposition starts to form and wavelength is too short to determine precisely. Since mechanical failure of materials, e.g. embrittlement of stainless steels, usually occurs at early stages of spinodal decomposition, estimation of composition amplitude is of the highest interest in engineering applications \cite{Brenner1982, Danoix2004}. Unfortunately, such a comparison is greatly impeded, since there is no method available yet to effectively analyze both continuum data from PFM and discrete data from APT. Therefore, a robust method to evaluate the composition amplitude in both simulations and experiments is urgently needed.

It is noteworthy that the most commonly adopted method for estimation of the composition amplitude in APT is the Langer-Bar-on-Miller (LBM) method \cite{Langer1975}, which can be also applied to the discrete data from MC simulations \cite{Hyde1995}. In the LBM method, the estimation of the composition amplitude is based on a frequency distribution histogram as shown in Figure 1(a) which stands for the probability of finding a certain composition in the analyzed volume of the alloy. In order to apply the LBM method, local composition should be estimated by sampling a certain number of atoms, a block, to calculate the average composition in that sampling block from the atom map as illustrated in Figure 1(a). In the LBM method, the frequency distribution histogram is represented by a continuous distribution function p(x) given by a sum of two Gaussian distributions with the same composition variance $\sigma$:

\begin{eqnarray} \label{eq:1}
p(x)= && \frac{1}{ (x_2-x_1)\sqrt{2\pi\sigma^2} } \biggl\{(x_2-x_0)exp \biggl[\frac{-(x-x_1)^2}{2\sigma^2}\biggr] \nonumber\\
         &&+(x_0-x_1)exp\biggl[\frac{-(x-x_2)^2}{2\sigma^2}\biggr] \biggr\},
\end{eqnarray}

\noindent where $x_1$ and $x_2$ are the compositions for the two Gaussian peaks, $x_0$ is the average composition in the analyzed volume of the sample. The experimental histogram thus is represented by 3 parameters  $x_1$, $x_2$ and $\sigma$, and the composition amplitude is $(x_2-x_1)$. However, the tails of the fitted Gaussian curves can sometimes extend beyond the physical limits of 0 and 1 at the composition axis (see Figure 1(a)). In fact, this may happen quite often when the sampling block for constructing concentration field is either comparable with or less than the half of the wavelength. Therefore, it artificially brings uncertainties into the analysis, and the fitted Gaussian distribution functions in the LBM method lose all statistical meaning. More importantly, when the tails extend to the unphysical region with the composition value below 0 or above 1, there is no guarantee that Eq.~(\ref{eq:1}) is consistent with the average composition of the alloy which would require that the integral from 0 to 1 must be unity. It has been suggested that some manipulations, like rescaling and constraining the data over the whole composition range, may be considered in order to eliminate that the tail exceeds the composition limits \cite{Hetherington1991,Miller2000}. Obviously, the accuracy of the LBM analysis cannot be guaranteed by taking such an expedient. 

\begin{figure}
\includegraphics{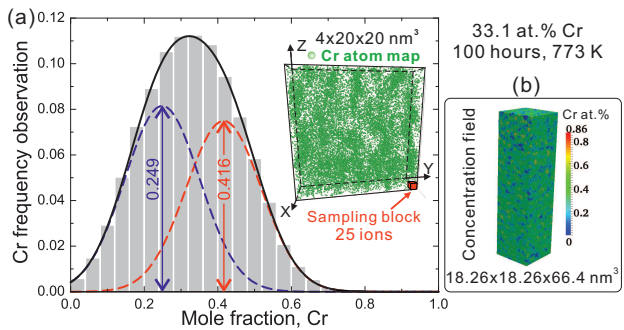}
\caption{\label{fig:1} (a) The frequency distribution histogram of the sample volume with 33.1 at.\% Cr isothermally aged at 773 K for 100 hours, constructed by using the block of 25 ions from the Cr atom map, which is partially shown on the right; (b) Constructed concentration field from the APT data. }
\end{figure}

One shall notice that the LBM method absolutely loses its applicability to the continuum data. For instance, the frequency distribution histograms that can be generated from PFM simulations differ quite much from the ones measured by APT, see Figure 2(a) to (c). Since the phase field model is based on a continuum model and the composition on each node of the simulation grid is obtained by solving the Cahn-Hilliard equation \cite{Cahn1958,Chen1998}, the obtained frequency distribution histogram for a real system is unlikely to show any value at 0 or 1 on the composition axis. Moreover, the frequency distribution histograms from phase field simulations usually exhibit much sharper changes than those obtained experimentally by APT with a bell curve shape shown in Figure 1(a)), and thus usually cannot be fitted by the Gaussian distribution functions of the LBM method. A histogram more similar to the experimental one may be obtained by considering the simulated value in a node as an average in a block of a given size and assume that the atoms within that block are randomly distributed which leads to a binomial distribution which may be approximated by a Gaussian if the block is not too small. 

Accordingly, there is a strong need to find a versatile method for evaluating the composition amplitude of spinodal decomposition for different techniques with discrete (APT or MC simulation) and continuum (PFM) information as the output. In this work, we propose such a new method, called Amplitude Density Spectrum (ADS), to estimate the composition amplitude in order to bridge different experimental and simulation tools for quantitative investigation of the spinodal microstructure evolution. Furthermore, the case study on the Fe-Cr alloy will indicate advantages of the ADS method in this work over the LBM method.

\section{Materials, experiments and simulations}
A binary Fe-Cr alloy was considered for testing the new method in the current work because of its importance to the investigation of spinodal decomposition in stainless steels \cite{Brenner1982,Lemoine1998,Park1995,Tane2003,Vitek1991}. Both APT and PFM were employed in the investigation.

\subsection{Atom probe tomography}
The Fe-Cr binary alloy with 37.8 at.\% Cr was produced by vacuum arc melting, and then homogenized at 1373 K for 2 hours under pure argon followed by quenching in brine. Afterwards, samples were annealed at 773 K for 100 hours, and then shaped into needle-like specimens for the standard two-stage electro-polishing. The analysis was performed under a local electrode atom probe (LEAP 3000X HRTM, Imago Scientific Instruments, USA) equipped with a reflectron for improved mass resolution. The ion detection efficiency is about 37 \%, and the experiments were made in voltage pulse mode (20 \% pulse fraction, 200 kHz, evaporation rate 1.5 \%) with a specimen temperature of 55 K. It should be noted that the analysis of the concentration in the analyzed volume in APT was found to be 33.1 at.\% Cr.

\subsection{Phase field simulation}
in order to verify the ADS method, a 3D phase field simulation on the Fe-35 at.\% Cr alloy was performed based on the Cahn-Hilliard equation:


\begin{widetext}
\begin{equation}
 \frac{\partial{x_{Cr}}}{\partial{t}}= \vec{\nabla} \cdot \biggl( x_{Cr}x_{Fe}\left(x_{Cr}M_{Fe}+x_{Fe}M_{Cr}\right)
\vec{\nabla} \biggl( \frac{\partial{G_m}}{\partial{x_{Cr}}}-\kappa\cdot\nabla^2x_{Cr} \biggr) \biggr). \label{eq:2}
\end{equation}
\end{widetext}

The free energy $G_m$ is obtained from the most recent thermodynamic modeling by Xiong et al. \cite{Xiong2010,Xiong2011}, while the parameters for describing atomic mobilities, $M_{Cr}$ and $M_{Fe}$, are taken from J\"{o}nsson \cite{Jonsson1995}. The simulation box is defined using   uniform grids, which is equivalent to $14\times14\times14$ $nm^3$. The composition gradient coefficient, $\kappa$, is defined as a function of the interatomic distance d (0.25nm), and regular solution parameter $\overline{\Omega}$(12225.15 J/mol at 773 K according to the improved thermodynamic modeling \cite{Xiong2011}), which equals to $\overline{\Omega}d^2/2$. The semi-implicit Fourier-spectral method \cite{Chen1998} is applied for simulation in a dimensionless form under periodic boundary conditions.

\begin{figure}
\includegraphics{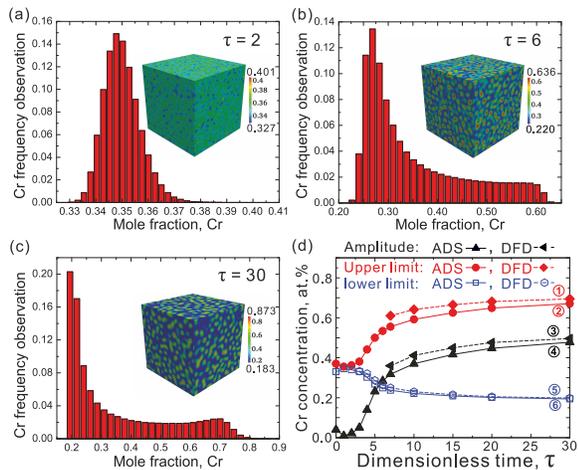}
\caption{\label{fig:2} Frequency distribution histogram and concentration field simulated by PFM at dimensionless time of (a) 2, (b) 6, and (c) 30. (d) Comparison of the analyzed results for composition amplitude and its upper and lower limit at different simulation times using DFD and ADS methods.}
\end{figure}

\section{Results and discussion}

\subsection{Analysis of APT data using the LBM and ADS methods}

Firstly, the LBM method was applied to estimate the composition amplitude for the prepared Fe-Cr sample. Since the original information from APT is the position of individual atoms as shown in Figure 1(a), the local composition is estimated at first in a certain size of block, which was chosen to be 25 ions in this work. As a result, the frequency distribution histogram for a given bin size, see Figure 1(a), can be achieved by statistically counting the numbers of concentration values within a certain composition range, i.e. composition limit of each bin in the frequency distribution histogram. Alternatively, a concentration field was constructed by using a sampling block with a volume of $0.83\times0.83\times0.83$ $nm^3$, if a cubic shape is considered for the sampling block with 25 ions. It is noteworthy that the frequency distribution histogram obtained in this way is almost identical to the previous one by counting the number of nodes of the concentration field with different concentration values in different concentration ranges. Using the LBM method, the frequency distribution histogram in Figure 1(a) can be fitted by two Gaussian distribution functions with the peaks at Cr content (mole fraction) of 0.249 and 0.416, respectively. As pointed out before, the frequency distribution histogram in Figure 1(a) has a tail exceeding the composition limit to negative values, which indicates that the Guassian function cannot represent the frequency distribution histogram satisfactory. Besides, one shall keep in mind that the outline of the histogram largely depends on its bin size. As a consequence, the choice of bin size is also a source of uncertainty in the LBM method.

\begin{figure}
\includegraphics{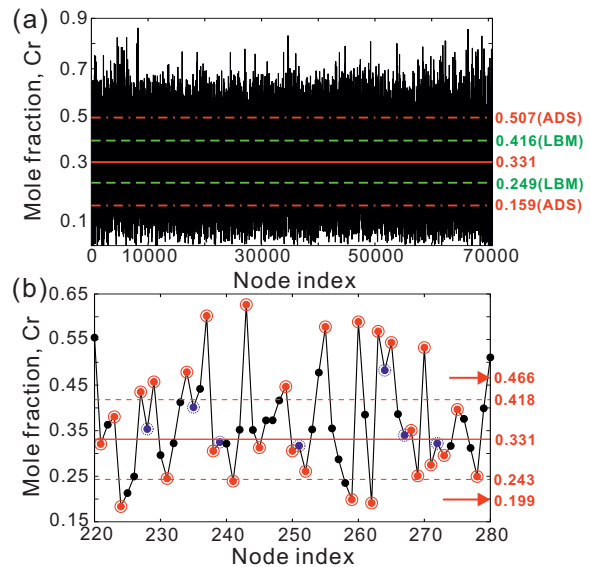}
\caption{\label{fig:3} (a) The ADS pattern constructed for the APT data of the sample with 33.1 at.\% Cr. The analyzed composition limits by the ADS method are denoted as the red chained lines, while the ones from the LBM method are indicated as the green dashed lines; (b) the magnified part of the ADS pattern. The dashed lines indicate the average composition limits (0.243 and 0.418) before the first refinement of the ADS method, while the composition limits pointed by arrows (0.199 and 0.466) are the ones after the first refinement of the ADS method. The blue dots marked in the dashed circles should be eliminated before the first refinement.}
\end{figure}

Based on the concentration field evaluated from APT, an ADS pattern could be directly constructed in Figure 3. In fact, the ADS pattern is a 1D plot of concentration values at each node in the concentration field following a spatial sequence over the whole analyzed volume. The overall ADS pattern normally contains much data points, and thus too dense to show clearly without large magnification, which can be found easily by comparing Figures 3(a) and (b). Therefore, it is necessary to establish a criterion to select peaks and troughs for evaluating the composition amplitude.

In this work, the evaluation of the composition amplitude is based on the following empirical steps, and eventually leads to results which can be verified further by PFM. The method seems to be robust and capable of determining small amplitudes. First, all of the peaks and troughs are selected to obtain the average concentration for the peaks and troughs. As indicated in Figure 3(b), the average Cr concentration for peaks is 0.418, while the one for troughs is 0.243, according to the ADS data in Figure 3(a). In order to achieve reasonable composition amplitude, some refinements are necessary. However, before refinement, we shall remove those peaks and troughs, as the ones marked in Figure 3(b) with blue dashed circles. The eliminated peaks are the ones lower than the average sample composition (0.331), and the excluded troughs are higher than the average. As the first step of the refinement, we only keep the peaks with the composition value higher than the above computed average peak composition (0.418), and the troughs lower than the computed average trough composition (0.243). As a consequence, the first refinement generates two average concentration values as the updated criteria for the second refinement: 0.466 for peaks and 0.199 for troughs as indicated in Figure 3(b). Based on these two new criteria, further refinements will be performed iteratively as in the first refinement. In the present case, the average composition values from each refinement for upper and lower limits are denoted as the red solid squares in Figure 4, respectively. Meanwhile, numbers of nodes at each step for refinement are plotted as the blue open circles. 

Furthermore, we found that the blue curves presenting the node numbers in terms of the refinement time are always varying smoothly. On the contrary, it is less common to get a smooth profile for the curve representing average Cr concentration. Analogous to defining deflections from a given baseline, e.g. onsets of transformations in a signal curve from differential thermal analysis, a criterion is set to determine the composition for upper and lower limits of the composition amplitude, which is simply making a tangent cross (see black dot A in Figure 4) of the blue curve for the number of nodes starting from the two ends as illustrated using the dashed lines in Figure 4. Afterwards, the composition limits can be read from the red star symbol B indicated on the composition curve, which has the same value of the vertical coordinate as point A. It is suggested that, in order to minimize the artificial manipulation in data analysis, the curve of the number of nodes will not be fitted by a polynomial to obtain a tangent cross. Instead, the tangent is directly obtained from the linear extension from the first two ending points in the blue curve for the number of nodes as illustrated in Figure 4.

As a result, the estimated composition amplitude of the sample volume with 33.1 at.\% Cr annealed at 773 K for 100 hours is 0.348, with the upper composition limit of 0.507 and lower limit of 0.159 as indicated in Figure 3(a). Apparently, the estimated composition amplitude from the ADS method (0.348) is much larger than the one by LBM (0.167). However, we shall notice that the average compositions of the peaks and troughs before refinements are 0.418 and 0.243 (see Figure 3(b)), which even show a larger composition difference, 0.175, than the evaluated composition amplitude from the LBM method. Obviously, large amount of gradient information has been included when computing the above two average composition values. Hence, the composition amplitude should be at least larger than 0.175. This also means that the LBM method has enormously underestimated the composition amplitude. 

Moreover, one may realize that the size of the sampling block will affect both the frequency distribution histogram and the concentration field based on the discrete data from APT. A reasonable sampling size should well represent the periodicity of the connected structure. However, the manipulation is difficult for 3D cases, and deserves further discussion in another work. On the contrary, the continuum data from phase field simulation can circumvent such a restriction due to sampling size. Therefore, the analysis of the phase field simulation results is performed. 

\begin{figure}
\includegraphics{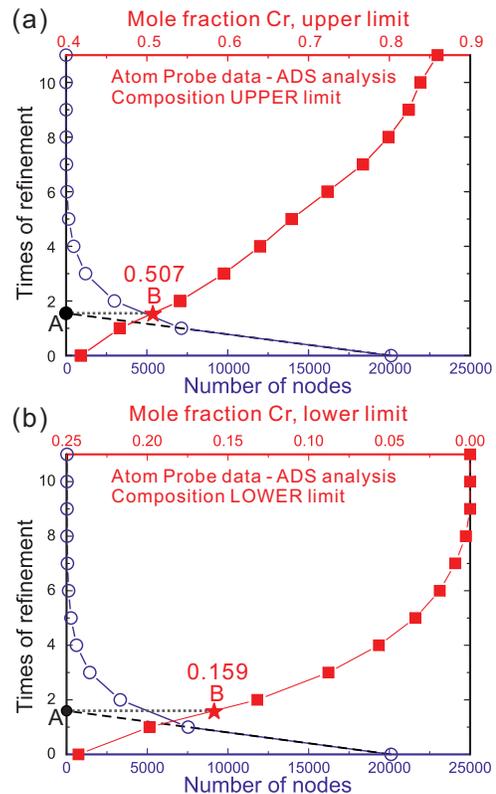}
\caption{\label{fig:4} The upper (a) and lower (b) limits of the composition amplitude analyzed by the ADS method. Blue open circles indicate number of nodes at each refinement step, while red solid squares are for calculated composition limits in each refinement. Points A and B have the same value on the vertical coordinate. }
\end{figure}

\section{Analysis of phase field simulation results}

It should be emphasized that the analysis of the phase field simulation result is desired not only to approve the validity of the now suggested ADS method but also to show its advantages of analyzing the continuum data over the LBM method.

In the present work, a typical feature of the frequency distribution histogram from PFM is shown in Figure 2. In the early stages, as pointed out before, the outline of the frequency is similar to the one read from APT with two peaks overlapping shown in Figure 2(a).  Figure 2(b) indicates a phase separation close to the medium stage. The peak on the Fe-rich side is pronounced, while the other peak is difficult to identify. However, when approaching to the later stages as shown in Figure 2(c), the peak on the Cr-rich side starts to be distinguishable. So far, the only method to define composition limits of amplitude is to determine the peak positions in frequency distribution histogram. If they are distinguishable as in the late stages (see Figure 2(c)), the composition amplitude will be the difference of the two composition limits corresponding to the peak position in the frequency distribution histogram. We named this as the direct frequency diagram (DFD) method.

Accordingly, the analyzed results by both DFD and ADS methods are summarized in Figure 2(d). The dashed lines marked with the odd numbers correspond to the DFD analysis once the peaks are pronounced enough for getting the composition limit of amplitude. Overall, the maximum difference of the composition amplitude between two different methods is 4.25 at.\% Cr at dimensionless time of 7, when the peak of frequency distribution histogram on the Cr-rich side starts to be distinguishable. Apparently, the overall agreement between these two methods is fairly good.
It should be noted that the definition of the composition amplitude in spinodal decomposition for real materials is blurred and depends on the determination method itself. An effective way to determine the composition amplitude should at least be capable to correctly and systemically demonstrate the variation of the composition amplitude along with the microstructure evolution. According to the analysis by the ADS method, the tendency of the variation for composition amplitudes is consistent with the one using the DFD analysis. Moreover, since the initial random noise set in the phase field simulation is 5.0 at.\% Cr, it is encouraging to point out that the analysis based on the ADS methods obtains a value of 4.01 at.\% Cr. Interestingly, the composition amplitude in the phase field simulation shows a minimum close to the dimensionless time of 1, which is consistent with the spinodal theory. Because the initial input state only includes the random noise and needs a certain incubation time to form the connected structure with certain wavelength under the thermodynamic and kinetic conditions \cite{Hillert1961}. The above analysis with PFM has successfully demonstrated the validity of the ADS methods.

The continuum PFM has various advantages in simulating the microstructural evolution. In order to integrate it with other experimental methods (e.g. APT) and simulation tools (e.g. MC, Molecular Dynamics), it is quite important to use a method, like ADS, to analyze the results containing both discrete and continuum information in the related studies.

\section{Conclusions}

In summary, a new method called ADS is developed in order to evaluate the composition amplitude from both discrete and continuum information. According to the analysis on the 3D phase field simulation results, the ADS method is accurate and effective. Based on the analysis of the APT data, the ADS method is approved being more accurate than the LBM method when applied to the early stages of decomposition in Fe-Cr alloys, at which the mechanical properties may drastically change, and thus needs sensitive measurement to determine the intensity of the spinodal decomposition. As a tool of microanalysis, the ADS method substantially bridges different experimental and simulation techniques for studying spinodal decomposition. 

\section*{Acknowledgments}

This work was performed within the VINN Excellence Center Hero-m, financed by VINNOVA, the Swedish Government Agency of Innovation Systems, Swedish Industry and KTH. The authors would like to thank Drs. Mattias Thuvander (Chalmers University of Technology), Peter Hedstr\"{o}m and Joakim Odqvist (KTH), and Prof. Emeritus Mats Hillert (KTH) for valuable discussions and help on experiments.


%


\begin{thebibliography}{10}%
\makeatletter
\providecommand \@ifxundefined [1]{%
 \ifx #1\undefined \expandafter \@firstoftwo
 \else \expandafter \@secondoftwo
\fi
}%
\providecommand \@ifnum [1]{%
 \ifnum #1\expandafter \@firstoftwo
 \else \expandafter \@secondoftwo
\fi
}%
\providecommand \enquote [1]{``#1''}%
\providecommand \bibnamefont  [1]{#1}%
\providecommand \bibfnamefont [1]{#1}%
\providecommand \citenamefont [1]{#1}%
\providecommand\href[0]{\@sanitize\@href}%
\providecommand\@href[1]{\endgroup\@@startlink{#1}\endgroup\@@href}%
\providecommand\@@href[1]{#1\@@endlink}%
\providecommand \@sanitize [0]{\begingroup\catcode`\&12\catcode`\#12\relax}%
\@ifxundefined \pdfoutput {\@firstoftwo}{%
 \@ifnum{\z@=\pdfoutput}{\@firstoftwo}{\@secondoftwo}%
}{%
 \providecommand\@@startlink[1]{\leavevmode}%
 \providecommand\@@endlink[0]{}%
}{%
 \providecommand\@@startlink[1]{%
  \leavevmode
  \pdfstartlink
   attr{/Border[0 0 1 ]/H/I/C[0 1 1]}%
   user{/Subtype/Link/A<</Type/Action/S/URI/URI(#1)>>}%
  \relax
 }%
 \providecommand\@@endlink[0]{\pdfendlink}%
}%
\providecommand \url  [0]{\begingroup\@sanitize \@url }%
\providecommand \@url [1]{\endgroup\@href {#1}{\urlprefix}}%
\providecommand \urlprefix [0]{URL }%
\providecommand \Eprint[0]{\href }%
\@ifxundefined \urlstyle {%
  \providecommand \doi [1]{doi:\discretionary{}{}{}#1}%
}{%
  \providecommand \doi [0]{doi:\discretionary{}{}{}\begingroup
  \urlstyle{rm}\Url }%
}%
\providecommand \doibase [0]{http://dx.doi.org/}%
\providecommand \Doi[1]{\href{\doibase#1}}%
\providecommand \selectlanguage [0]{\@gobble}%
\providecommand \bibinfo [0]{\@secondoftwo}%
\providecommand \bibfield [0]{\@secondoftwo}%
\providecommand \translation [1]{[#1]}%
\providecommand \BibitemOpen[0]{}%
\providecommand \bibitemStop [0]{}%
\providecommand \bibitemNoStop [0]{.\EOS\space}%
\providecommand \EOS [0]{\spacefactor3000\relax}%
\providecommand \BibitemShut [1]{\csname bibitem#1\endcsname}%
\bibitem{Marquis2010}%
  \BibitemOpen
  \bibfield{author}{%
  \bibinfo {author} {\bibfnamefont{E.~A.}\ \bibnamefont{Marquis}}\ and\
  \bibinfo {author} {\bibfnamefont{J.~M.}\ \bibnamefont{Hyde}},\ }%
  \bibfield{journal}{%
  \bibinfo {journal} {Mater. Sci. Eng. R.}\ }%
  \textbf{\bibinfo {volume} {69}},\ \bibinfo {pages} {37} (\bibinfo {year}
  {2010})\BibitemShut{NoStop}%
\bibitem{Miller1996}%
  \BibitemOpen
  \bibfield{author}{%
  \bibinfo {author} {\bibfnamefont{M.K.}\ \bibnamefont{Miller}},\ }%
  \emph{\bibinfo {title} {Atom probe field ion microscopy}}\ (\bibinfo
  {publisher} {Clarendon Press},\ \bibinfo {year} {1996})\ ISBN \bibinfo {isbn}
  {9780198513872}\BibitemShut{NoStop}%
\bibitem{Seidman2007}%
  \BibitemOpen
  \bibfield{author}{%
  \bibinfo {author} {\bibfnamefont{D.~N.}\ \bibnamefont{Seidman}},\ }%
  \bibfield{journal}{%
  \bibinfo {journal} {Annu. Rev. Mater. Res.}\ }%
  \textbf{\bibinfo {volume} {37}},\ \bibinfo {pages} {127} (\bibinfo {year}
  {2007})\BibitemShut{NoStop}%
\bibitem{Hillert1961}%
  \BibitemOpen
  \bibfield{author}{%
  \bibinfo {author} {\bibfnamefont{M.}~\bibnamefont{Hillert}},\ }%
  \bibfield{journal}{%
  \bibinfo {journal} {Acta Metall.}\ }%
  \textbf{\bibinfo {volume} {9}},\ \bibinfo {pages} {525} (\bibinfo {year}
  {1961})\BibitemShut{NoStop}%
\bibitem{Cahn1958}%
  \BibitemOpen
  \bibfield{author}{%
  \bibinfo {author} {\bibfnamefont{J.~W.}\ \bibnamefont{Cahn}}\ and\ \bibinfo
  {author} {\bibfnamefont{J.~E.}\ \bibnamefont{Hilliard}},\ }%
  \bibfield{journal}{%
  \bibinfo {journal} {J. Chem. Phys}\ }%
  \textbf{\bibinfo {volume} {28}},\ \bibinfo {pages} {258} (\bibinfo {year}
  {1958})\BibitemShut{NoStop}%
\bibitem{Brenner1982}%
  \BibitemOpen
  \bibfield{author}{%
  \bibinfo {author} {\bibfnamefont{S.S.}\ \bibnamefont{Brenner}}, \bibinfo
  {author} {\bibfnamefont{M.K.}\ \bibnamefont{Miller}},\ and\ \bibinfo {author}
  {\bibfnamefont{W.A.}\ \bibnamefont{Soffa}},\ }%
  \bibfield{journal}{%
  \bibinfo {journal} {Scr. Metall.}\ }%
  \textbf{\bibinfo {volume} {16}},\ \bibinfo {pages} {831} (\bibinfo {year}
  {1982})\BibitemShut{NoStop}%
\bibitem{Danoix2004}%
  \BibitemOpen
  \bibfield{author}{%
  \bibinfo {author} {\bibfnamefont{F.}~\bibnamefont{Danoix}}, \bibinfo {author}
  {\bibfnamefont{P.}~\bibnamefont{Auger}},\ and\ \bibinfo {author}
  {\bibfnamefont{D.}~\bibnamefont{Blavette}},\ }%
  \bibfield{journal}{%
  \bibinfo {journal} {Microsc. Microanal.}\ }%
  \textbf{\bibinfo {volume} {10}},\ \bibinfo {pages} {349} (\bibinfo {year}
  {2004})\BibitemShut{NoStop}%
\bibitem{Langer1975}%
  \BibitemOpen
  \bibfield{author}{%
  \bibinfo {author} {\bibfnamefont{J.~S.}\ \bibnamefont{Langer}}, \bibinfo
  {author} {\bibfnamefont{M.}~\bibnamefont{Bar-on}},\ and\ \bibinfo {author}
  {\bibfnamefont{H.~D.}\ \bibnamefont{Miller}},\ }%
  \bibfield{journal}{%
  \bibinfo {journal} {Phys. Rev. A}\ }%
  \textbf{\bibinfo {volume} {11}},\ \bibinfo {pages} {1417} (\bibinfo {year}
  {1975})\BibitemShut{NoStop}%
\bibitem{Hyde1995}%
  \BibitemOpen
  \bibfield{author}{%
  \bibinfo {author} {\bibfnamefont{J.M.}\ \bibnamefont{Hyde}}, \bibinfo
  {author} {\bibfnamefont{M.K.}\ \bibnamefont{Miller}}, \bibinfo {author}
  {\bibfnamefont{M.G.}\ \bibnamefont{Hetherington}}, \bibinfo {author}
  {\bibfnamefont{A.}~\bibnamefont{Cerezo}}, \bibinfo {author}
  {\bibfnamefont{G.D.W.}\ \bibnamefont{Smith}},\ and\ \bibinfo {author}
  {\bibfnamefont{C.M.}\ \bibnamefont{Elliott}},\ }%
  \bibfield{journal}{%
  \bibinfo {journal} {Acta Mater.}\ }%
  \textbf{\bibinfo {volume} {43}},\ \bibinfo {pages} {3403} (\bibinfo {year}
  {1995})\BibitemShut{NoStop}%
\bibitem{Hetherington1991}%
  \BibitemOpen
  \bibfield{author}{%
  \bibinfo {author} {\bibfnamefont{M.~G.}\ \bibnamefont{Hetherington}},
  \bibinfo {author} {\bibfnamefont{J.~M.}\ \bibnamefont{Hyde}}, \bibinfo
  {author} {\bibfnamefont{M.~K.}\ \bibnamefont{Miller}},\ and\ \bibinfo
  {author} {\bibfnamefont{G.~D.~W.}\ \bibnamefont{Smith}},\ }%
  \bibfield{journal}{%
  \bibinfo {journal} {Surf. Sci.}\ }%
  \textbf{\bibinfo {volume} {246}},\ \bibinfo {pages} {304} (\bibinfo {year}
  {1991})\BibitemShut{NoStop}%
\bibitem{Miller2000}%
  \BibitemOpen
  \bibfield{author}{%
  \bibinfo {author} {\bibfnamefont{M.K.}\ \bibnamefont{Miller}},\ }%
  \emph{\bibinfo {title} {Atom Probe Tomography: Analysis at the Atomic
  Level}},\ \bibinfo {edition} {1st}\ ed.\ (\bibinfo {publisher} {Springer},\
  \bibinfo {year} {2000})\BibitemShut{NoStop}%
\bibitem{Chen1998}%
  \BibitemOpen
  \bibfield{author}{%
  \bibinfo {author} {\bibfnamefont{L.~Q.}\ \bibnamefont{Chen}}\ and\ \bibinfo
  {author} {\bibfnamefont{J.}~\bibnamefont{Shen}},\ }%
  \bibfield{journal}{%
  \bibinfo {journal} {Comput. Phys. Commun.}\ }%
  \textbf{\bibinfo {volume} {108}},\ \bibinfo {pages} {147} (\bibinfo {year}
  {1998})\BibitemShut{NoStop}%
\bibitem{Lemoine1998}%
  \BibitemOpen
  \bibfield{author}{%
  \bibinfo {author} {\bibfnamefont{C.}~\bibnamefont{Lemoine}}, \bibinfo
  {author} {\bibfnamefont{A.}~\bibnamefont{Fnidiki}}, \bibinfo {author}
  {\bibfnamefont{J.}~\bibnamefont{Teillet}}, \bibinfo {author}
  {\bibfnamefont{M.}~\bibnamefont{Hedin}},\ and\ \bibinfo {author}
  {\bibfnamefont{F.}~\bibnamefont{Danoix}},\ }%
  \bibfield{journal}{%
  \bibinfo {journal} {Scr. Mater.}\ }%
  \textbf{\bibinfo {volume} {39}},\ \bibinfo {pages} {61} (\bibinfo {year}
  {1998})\BibitemShut{NoStop}%
\bibitem{Park1995}%
  \BibitemOpen
  \bibfield{author}{%
  \bibinfo {author} {\bibfnamefont{J.~S.}\ \bibnamefont{Park}}\ and\ \bibinfo
  {author} {\bibfnamefont{Y.~K.}\ \bibnamefont{Yoon}},\ }%
  \bibfield{journal}{%
  \bibinfo {journal} {Scr. Metall. Mater.}\ }%
  \textbf{\bibinfo {volume} {32}},\ \bibinfo {pages} {1163} (\bibinfo {year}
  {1995})\BibitemShut{NoStop}%
\bibitem{Tane2003}%
  \BibitemOpen
  \bibfield{author}{%
  \bibinfo {author} {\bibfnamefont{M.}~\bibnamefont{Tane}}, \bibinfo {author}
  {\bibfnamefont{T.}~\bibnamefont{Ichitsubo}}, \bibinfo {author}
  {\bibfnamefont{H.}~\bibnamefont{Ogi}},\ and\ \bibinfo {author}
  {\bibfnamefont{M.}~\bibnamefont{Hirao}},\ }%
  \bibfield{journal}{%
  \bibinfo {journal} {Scr. Mater.}\ }%
  \textbf{\bibinfo {volume} {48}},\ \bibinfo {pages} {229} (\bibinfo {year}
  {2003})\BibitemShut{NoStop}%
\bibitem{Vitek1991}%
  \BibitemOpen
  \bibfield{author}{%
  \bibinfo {author} {\bibfnamefont{J.~M.}\ \bibnamefont{Vitek}}, \bibinfo
  {author} {\bibfnamefont{S.~A.}\ \bibnamefont{David}}, \bibinfo {author}
  {\bibfnamefont{D.~J.}\ \bibnamefont{Alexander}}, \bibinfo {author}
  {\bibfnamefont{J.~R.}\ \bibnamefont{Keiser}},\ and\ \bibinfo {author}
  {\bibfnamefont{R.~K.}\ \bibnamefont{Nanstad}},\ }%
  \bibfield{journal}{%
  \bibinfo {journal} {Acta Metall. Mater.}\ }%
  \textbf{\bibinfo {volume} {39}},\ \bibinfo {pages} {503} (\bibinfo {year}
  {1991})\BibitemShut{NoStop}%
\bibitem{Xiong2010}%
  \BibitemOpen
  \bibfield{author}{%
  \bibinfo {author} {\bibfnamefont{W.}~\bibnamefont{Xiong}}, \bibinfo {author}
  {\bibfnamefont{M.}~\bibnamefont{Selleby}}, \bibinfo {author}
  {\bibfnamefont{Q.}~\bibnamefont{Chen}}, \bibinfo {author}
  {\bibfnamefont{J.}~\bibnamefont{Odqvist}},\ and\ \bibinfo {author}
  {\bibfnamefont{Y.}~\bibnamefont{Du}},\ }%
  \bibfield{journal}{%
  \bibinfo {journal} {Crit. Rev. Solid State Mater. Sci.}\ }%
  \textbf{\bibinfo {volume} {35}},\ \bibinfo {pages} {125} (\bibinfo {year}
  {2010})\BibitemShut{NoStop}%
\bibitem{Xiong2011}%
  \BibitemOpen
  \bibfield{author}{%
  \bibinfo {author} {\bibfnamefont{W.}~\bibnamefont{Xiong}}, \bibinfo {author}
  {\bibfnamefont{J.}~\bibnamefont{Odqvist}}, \bibinfo {author}
  {\bibfnamefont{P.}~\bibnamefont{Hedstrom}}, \bibinfo {author}
  {\bibfnamefont{M.}~\bibnamefont{Selleby}}, \bibinfo {author}
  {\bibfnamefont{M.}~\bibnamefont{Thuvander}},\ and\ \bibinfo {author}
  {\bibfnamefont{Q.}~\bibnamefont{Chen}},\ }%
  \bibfield{journal}{%
  \bibinfo {journal} {CALPHAD}\ }%
  \textbf{\bibinfo {volume} {35}},\ \bibinfo {pages} {355} (\bibinfo {year}
  {2011})\BibitemShut{NoStop}%
\bibitem{Jonsson1995}%
  \BibitemOpen
  \bibfield{author}{%
  \bibinfo {author} {\bibfnamefont{B.}~\bibnamefont{Jonsson}},\ }%
  \bibfield{journal}{%
  \bibinfo {journal} {ISIJ Int.}\ }%
  \textbf{\bibinfo {volume} {35}},\ \bibinfo {pages} {1415} (\bibinfo {year}
  {1995})\BibitemShut{NoStop}%
\end{thebibliography}
\end{document}